\documentstyle[12pt]{article}

\font\bdi=cmmib10 at 12 pt

\def\bi#1{\hbox{\bdi #1\/}}
\def\xb{\bi{x}}

\def\sigmab{\bi{\char'33}}

\newcommand{\D}{\hat D}
\newcommand{\tP}{\tilde P}
\newcommand{\diag}{\mathop{\rm diag}\nolimits}
\newcommand{\sumjk}{\sum_{\scriptstyle j,k \atop \scriptstyle i\ne j\ne k\ne i}}
\newcommand{\sumij}{\sum_{\scriptstyle i,j \atop \scriptstyle i\ne j}}
\def\case#1#2{{\textstyle{#1\over #2}}}

\begin{document}

\baselineskip=22pt plus 1pt minus 1pt
%
%
\begin{center}
{\Large \bf Three-body generalization of the Sutherland model}

{\Large \bf with internal degrees of freedom} 

\bigskip\bigskip
{C. Quesne \footnote{Directeur de recherches FNRS; E-mail: cquesne@ulb.ac.be}}

\bigskip
{\it Physique Nucl\'eaire Th\'eorique et Physique Math\'ematique, 
Universit\'e Libre de Bruxelles, Campus de la Plaine CP229, Boulevard du 
Triomphe, B-1050 Brussels, Belgium}
\end{center}

\bigskip\bigskip\bigskip\bigskip
%
%

\begin{abstract}

\noindent
A generalized spin Sutherland model including a three-body potential is proposed.
The problem is analyzed in terms of three first-order differential-difference
operators, obtained by combining SUSYQM supercharges with the elements of the
dihedral group~$D_6$. Three alternative commuting operators are also introduced.

\end{abstract}

\vspace{5cm}

\clearpage
%
%
In recent years, the Sutherland one-dimensional $N$-particle
model~\cite{sutherland} and its rational limit, the Calogero
model~\cite{calogero}, have received considerable attention in the literature
because they are relevant to several important physical problems (for a list of
references, see e.g.~\cite{cq}).\par
%
%
The Sutherland problem can be analyzed in terms of a set of $N$ commuting
first-order differential-difference operators~\cite{poly}, related to the root
system of the ${\cal A}_{N-1}$ algebra~\cite{perelomov} and known in the
mathematical literature as Dunkl operators~\cite{dunkl}. Use of the latter leads to
a Hamiltonian with exchange terms, connected with an extension of the model for
particles with internal degrees of freedom, referred to as the spin Sutherland
problem~\cite{ha, bernard}.\par
%
%
A similar type of approach can be employed~\cite{buchstaber} for other integrable
models related to root systems of Lie algebras~\cite{perelomov}.\par
%
%
In the present letter, we shall deal with a generalized Sutherland three-particle
problem including an extra three-body trigonometric potential. Such a problem is
related to the exceptional Lie algebra $G_2$, whose Weyl group, of order~12, is
the dihedral group $D_6$~\cite{perelomov}. In contrast with the approaches used
elsewhere~\cite{poly,dunkl,ha,bernard,buchstaber}, our starting point will be an
analysis of the problem in supersymmetric quantum mechanics (SUSYQM), thereby
emphasizing the link between Dunkl operators and SUSYQM.\par
%
%
Let us consider a system of three particles on a circle of length~$\pi/a$
interacting via long-range two- and three-body potentials. Its Hamiltonian is
defined by 
\begin{equation}
  H = - \sum_{i=1}^3 \partial_i^2  
       + g a^2 \sum_{\scriptstyle i,j=1 \atop \scriptstyle i\ne j}^3 \csc^2
       \left(a(x_i-x_j)\right)
       + 3f a^2  \sum_{\scriptstyle i,j,k=1 \atop \scriptstyle i\ne j\ne k\ne i}^3 
       \csc^2  \left(a(x_i+x_j-2x_k)\right), \label{eq:H}
\end{equation}
where $x_i$, $i=1$, 2,~3, $0\le x_i\le \pi/a$, denote the particle coordinates,
$\partial_i \equiv \partial/\partial x_i$, and $g$, $f$ are assumed not to vanish
simultaneously and to be such that $g > -1/4$, $f > -1/4$. In the case where $g\ne
0$ and $f = 0$, Hamiltonian~(\ref{eq:H}) reduces to the Sutherland
Hamiltonian~\cite{sutherland}.\par
%
%
Throughout this paper, we shall use the notations $x_{ij} \equiv
x_i - x_j$, $i\ne j$, and $y_{ij} \equiv x_i + x_j - 2x_k$, $i\ne j\ne k\ne i$, where
in the latter, index~$k$ is suppressed as it is entirely determined by $i$
and~$j$. Except where otherwise stated, we shall assume that the particles are
distinguishable. In the case of indistinguishable particles, an additional
symmetry requirement has to be imposed on the wave functions.\par
%
%
For distinguishable particles, the unnormalized ground-state wave function of
Hamiltonian~(\ref{eq:H}), is given by $\psi_0(\xb) = \prod_{i\ne j}
\left|\sin(a x_{ij})\right|^{\kappa} \left|\sin(a y_{ij})\right|^{\lambda}$, and
corresponds to the eigenvalue $E_0 = 8 a^2 (\kappa^2 + 3 \kappa \lambda + 3
\lambda^2)$, where 
\begin{equation}
  \kappa \equiv \left\{\begin{array}{ll}
                                     \case{1}{2} (1 + \sqrt{1 + 4g}) & \mbox{if $g\ne 0$}
                                          \\[0.1cm]
                                     0                                              & \mbox{if $g=0$}  
                                   \end{array}\right., \qquad
  \lambda \equiv \left\{\begin{array}{ll}
                                     \case{1}{2} (1 + \sqrt{1 + 4f}) & \mbox{if $f\ne 0$}
                                          \\[0.1cm]
                                     0                                              & \mbox{if $f=0$}  
                                     \end{array}\right.,
  \label{eq:kappa-lambda}
\end{equation}
or, equivalently, $g = \kappa (\kappa - 1)$, $f = \lambda (\lambda - 1)$. The proof
of this result is based upon the trigonometric identities
\begin{eqnarray}
  \sum_{\scriptstyle i,j,k \atop \scriptstyle i\ne j\ne k\ne i} \cot(a x_{ij})
       \cot(a x_{jk}) & = &  \sum_{\scriptstyle i,j,k \atop \scriptstyle i\ne j\ne k\ne 
       i} \cot(a y_{ij}) \cot(a y_{jk}) = 2, \nonumber \\
   \sum_{\scriptstyle i,j,k \atop \scriptstyle i\ne j\ne k\ne i} \cot(a x_{ij})
       \cot(a y_{jk}) & = & 4.        \label{eq:trig2}  
\end{eqnarray}
\par
%
%
The three-particle Hamiltonian~(\ref{eq:H}) can be alternatively considered as that
of a particle in three-dimensional space. By using the Andrianov {\em et
al} generalization of SUSYQM for multidimensional Hamiltonians~\cite{andrianov},
$H - E_0$ can therefore be regarded as the $H^{(0)}$ component of a supersymmetric
Hamiltonian $\hat H = \diag\left(H^{(0)}, H^{(1)}, H^{(2)}, H^{(3)}\right)$ with
supercharge operators $\hat Q^+$, $\hat Q^- = \left(\hat Q^+\right)^{\dagger}$. The
matrix elements of the latter can be expressed in terms of six differential
operators
\begin{equation}
  Q^{\pm}_i = \mp \partial_i - \kappa a \sum_{j\ne i} \cot(a x_{ij}) 
  - \lambda a \left(\sum_{j\ne i} \cot(a y_{ij}) - \sumjk \cot(a y_{jk})\right),
  \qquad i = 1,2,3,      \label{eq:charges}
\end{equation}
which are obtained from the ground-state wave function by using the recipe
$Q_i^{\pm} = \mp \partial_i + \partial_i \chi(\xb)$, where $\chi(\xb) = - \ln
\psi_0(\xb)$. In terms of such operators, $H^{(0)} = Q^+_i Q^-_i$ while $H^{(3)} =
Q^-_i Q^+_i$. Apart from some additive constant, the latter turns out to be given
by~(\ref{eq:H}) with $g = \kappa (\kappa - 1)$, $f = \lambda (\lambda - 1)$ replaced
by $g = \kappa (\kappa + 1)$, $f = \lambda (\lambda + 1)$, respectively.\par
%
%
Let us now transform the supercharge operators $Q^-_i$ of Eq.~(\ref{eq:charges})
into some differential-difference operators
\begin{equation}
  D_i = \partial_i - \kappa a \sum_{j\ne i} \cot(a x_{ij}) K_{ij} 
  - \lambda a \left(\sum_{j\ne i} \cot(a y_{ij}) L_{ij} - \sumjk \cot(a y_{jk}) L_{jk}
  \right),      \label{eq:D}
\end{equation}
by inserting some finite group elements $K_{ij}$ and $L_{ij} \equiv K_{ij} I_r$. Here
$K_{ij}$ are particle permutation operators, while $I_r$ is the inversion operator in
relative-coordinate space. In the centre-of-mass coordinate system to be
used in the remainder of this paper, they satisfy the relations
\begin{eqnarray}
  K_{ij} & = & K_{ji} = K_{ij}^{\dagger}, \qquad K_{ij}^2 = 1, \qquad K_{ij} K_{jk} =
          K_{jk} K_{ki} = K_{ki} K_{ij}, \nonumber \\
  K_{ij} I_r & = & I_r K_{ij}, \qquad I_r = I_r^{\dagger}, \quad I_r^2 = 1,      
          \label{eq:K-Ir1} \\
  K_{ij} x_j & = & x_i K_{ij}, \quad K_{ij} x_k = x_k K_{ij}, \quad I_r x_i = - x_i
          I_r,       \label{eq:K-Ir2}
\end{eqnarray}
for all $i\ne j\ne k\ne i$. The operators 1, $K_{ij}$, $K_{ijk} \equiv K_{ij} K_{jk}$,
$I_r$, $L_{ij}$, and $L_{ijk} \equiv K_{ijk} I_r$, where $i$, $j$, $k$ run over the set
\{1, 2, 3\}, are the 12 elements of the dihedral group~$D_6$~\cite{hamermesh}.\par
%
%
From their definition and Eqs.~(\ref{eq:K-Ir1}),~(\ref{eq:K-Ir2}), it is obvious that
the differential-difference operators~$D_i$ are both antihermitian and
$D_6$-covariant, i.e., $D_i^{\dagger} = - D_i$, $K_{ij} D_j = D_i K_{ij}$,
$K_{ij}D_k = D_k K_{ij}$, and $I_r D_i = - D_i I_r$, for all $i\ne j\ne k\ne i$.
After some straightforward, although rather lengthy, calculations
using again the tri\-gonometric identities~(\ref{eq:trig2}), one obtains that their
commutators are given by
\begin{equation}
  \left[D_i, D_j\right] = - a^2 \left(\kappa^2 + 3 \lambda^2 - 4 \kappa \lambda I_r
  \right) \sum_{k\ne i,j} \left(K_{ijk} - K_{ikj}\right), \qquad i\ne j, 
  \label{eq:D-com}
\end{equation}
and that
\begin{eqnarray}
  -\sum_i D_i^2 & = & - \sum_i \partial_i^2 + a^2 \sumij \csc^2(a x_{ij}) \kappa
       (\kappa - K_{ij}) + 3 a^2 \sumij \csc^2(a y_{ij}) \lambda (\lambda - L_{ij})
       \nonumber\\
  & & - 6 a^2 \left(\kappa^2 + 3 \lambda^2\right) - a^2 \left(\kappa^2 + 
       3 \lambda^2 + 12 \kappa \lambda I_r\right) \left(K_{123} + K_{132}\right).
       \label{eq:D-square}     
\end{eqnarray}
\par
%
%
From Eq.~(\ref{eq:D-com}), it is clear that the operators~$D_i$ do not commute
among themselves, except in the $a \to 0$ limit, i.e., in the rational case
considered many years ago by Wolfes~\cite{wolfes}, and by Calogero and
Marchioro~\cite{marchioro}. Furthermore, Eq.~(\ref{eq:D-square}) shows that the
generalized Hamiltonian with exchange terms
\begin{equation}
  H_{exch} = - \sum_i \partial_i^2 + a^2 \sumij \csc^2(a x_{ij}) \kappa
  (\kappa - K_{ij}) + 3 a^2 \sumij \csc^2(a y_{ij}) \lambda (\lambda - L_{ij})
  \label{eq:Hexch}
\end{equation}
only differs by some exchange operators from the Hamiltonian $\sum_i \pi_i^2$,
written in terms of the generalized momenta $\pi_i = \pi_i^{\dagger} = - i D_i$. In
those subspaces of Hilbert space wherein $\left(K_{ij}, L_{ij}\right) = (1,1)$,
$(1,-1)$, $(-1,1)$, or~$(-1,-1)$, $H_{exch}$ reduces to
Hamiltonian~(\ref{eq:H}) corresponding to $(g,f) = (\kappa (\kappa-1), \lambda
(\lambda-1))$, $(\kappa (\kappa-1), \lambda (\lambda+1))$, $(\kappa (\kappa+1),
\lambda (\lambda-1))$, or $(\kappa (\kappa+1), \lambda (\lambda+1))$,
respectively.\par
%
%
As in the case of the Sutherland problem~\cite{bernard}, we can try to reformulate
the present one in terms of some commuting, albeit non-covariant, differential
operators~$\D_i$. Let
\begin{equation}
  \D_i = D_i + i \kappa a \sum_{j\ne i} \alpha_{ij} K_{ij} + i \lambda a \left(
  \sum_{j\ne i} \beta_{ij} L_{ij} - \sumjk \beta_{jk} L_{jk}\right),    \label{eq:hatD}
\end{equation}
where $\alpha_{ij}$ and $\beta_{ij}$ are some real constants. With such a choice,
the transformed operators remain antihermitian, i.e., $\D_i^{\dagger} = - \D_i$.
We shall assume in addition that $\alpha_{ji} = - \alpha_{ij}$ and $\beta_{ji} =
\beta_{ij}$. This assumption is justified by the fact that for~$\lambda =
0$~\cite{bernard}, the operators~(\ref{eq:hatD}) with $\alpha_{ij} = - \alpha_{ji} =
- 1$, $i<j$, do fulfil the required property $\left[\D_i, \D_j\right] = 0$.\par
%
%
From definition~(\ref{eq:hatD}) and the $D_6$-covariance of~$D_i$,
the new operators~$\D_i$ have the following transformation properties
under~$D_6$:
\begin{eqnarray}
  K_{ij} \D_j - \D_i K_{ij} & = & - i \kappa a \left(2 \alpha_{ij} + \sum_{k\ne i,j}
       \left(\alpha_{ik} - \alpha_{jk}\right) K_{ijk}\right) \nonumber \\
  & & - i \lambda a \sum_{k\ne i,j} \left(\beta_{ik} - \beta_{jk}\right) I_r 
       \left(K_{ijk} + 2 K_{ikj}\right), \qquad i\ne j,      \label{eq:hatD-cov1} \\
  \left[K_{ij}, \D_k\right] & = & i a \left[\kappa \left(\alpha_{ik} - \alpha_{jk}
       \right) - \lambda \left(\beta_{ik} - \beta_{jk}\right) I_r\right] \left(K_{ijk}
       - K_{ikj}\right), \nonumber \\
  & & i\ne j\ne k\ne i,       \label{eq:hatD-cov2} \\
  \left\{I_r, \D_i\right\} & = & 2i \kappa a \sum_{j\ne i} \alpha_{ij} L_{ij}
       + 2i \lambda a \left(\sum_{j\ne i} \beta_{ij} K_{ij} - \sumjk \beta_{jk} K_{jk}
       \right).          \label{eq:hatD-cov3}
\end{eqnarray}
Hence, they can only be covariant provided $\alpha_{ij} = \beta_{ij} = 0$, i.e., when
they coincide with the $D_i$'s.\par
%
%
By using various properties of the operators~$D_i$, as well as the relation $\sum_i
D_i = 0$ valid in the centre-of-mass coordinate system, it is straightforward to
show that if the constants
$\alpha_{ij}$, $\beta_{ij}$ fulfil the three relations
\begin{equation}
  \alpha_{12} \alpha_{23} + \alpha_{23} \alpha_{31} + \alpha_{31} \alpha_{12}
  =  -1, \qquad \beta_{12}\beta_{23} + \beta_{23} \beta_{31} + \beta_{31} 
  \beta_{12} = -1,        \label{eq:cond1} 
\end{equation}
\begin{equation}
  (\alpha_{23} - \alpha_{31}) \beta_{12} + (\alpha_{31} - \alpha_{12}) \beta_{23}
  + (\alpha_{12} - \alpha_{23}) \beta_{31} = 4,      \label{eq:cond2}
\end{equation}
then the operators $\D_i$ commute among themselves, and $H_{exch}$ only differs
by some additive constant from the Hamiltonian $\sum_i \hat{\pi}_i^2$, where
$\hat{\pi}_i = \hat{\pi}_i^{\dagger} = - i \D_i$. Furthermore, this additive constant
can be set equal to zero by normalizing $\alpha_{ij}$ and~$\beta_{ij}$ in such a
way that the relations
\begin{equation}
  \alpha_{12}^2 + \alpha_{23}^2 + \alpha_{31}^2 = 3, \qquad 
  \beta_{12}^2 + \beta_{23}^2 + \beta_{31}^2 = 3,      \label{eq:norm}
\end{equation}
are satisfied.\par
%
%
The constants $\alpha_{ij} = - \alpha_{ji} = -1$, $i<j$, used in Ref.~\cite{bernard},
do fulfil the first relation in both Eqs.~(\ref{eq:cond1}) and~(\ref{eq:norm}). It
should be stressed however that such equations admit other solutions too. It is
then easy to prove that the remaining three relations in
(\ref{eq:cond1})--(\ref{eq:norm}) admit four, and only four, solutions for the
$\beta_{ij}$'s, compatible with this choice for the~$\alpha_{ij}$'s: $(\beta_{12},
\beta_{23}, \beta_{31}) = (-1,1,1)$, $(-1,1,-1)$, $(- 5/3, 1/3, 1/3)$, and $(- 1/3,
5/3, - 1/3)$.\par
%
%
A relation can be established between $H_{exch}$ and a Hamiltonian ${\cal
H}^{(\kappa,\lambda)}$ describing a one-dimensional system of three particles
with $SU(n)$ ``spins'' (or colours in particle physics language), interacting via
spin-dependent two and three-body potentials,
\begin{equation}
  {\cal H}^{(\kappa,\lambda)} = - \sum_i \partial_i^2 + a^2 \sumij \csc^2(a x_{ij})
  \kappa (\kappa - P_{ij}) + 3 a^2 \sumij \csc^2(a y_{ij}) \lambda (\lambda -
  \tilde P_{ij}).      \label{eq:spinH}
\end{equation}
Here each particle is assumed to carry a spin with $n$ possible values, and
$P_{ij}$, $\tilde P_{ij} \equiv P_{ij} \tilde P$ are some operators acting only in spin
space. The operator~$P_{ij}$ is defined as the operator permuting the $i$th and
$j$th spins, while $\tP$ is a permutation-invariant and involutive operator, i.e.,
$\tP \sigma_i = \sigma_i^* \tP$, for some $\sigma^*_i$ such that
$P_{jk} \sigma_i^* = \sigma_i^* P_{jk}$ for all $i$, $j$,~$k$, and $\sigma_i^{**} =
\sigma_i$. For $SU(2)$ spins for instance, $\sigma_i = \pm 1/2$, $P_{ij} =
(\sigma^a_i \sigma^a_j + 1)/2$, where
$\sigma^a$, $a=1$, 2,~3, denote the Pauli matrices, $\tP$ may be taken as 1 or
$\sigma^1_1 \sigma^1_2 \sigma^1_3$, and accordingly $\sigma_i^* =
\sigma_i$ or~$-\sigma_i$. The operators $P_{ij}$ and~$\tP$ satisfy
relations similar to those fulfilled by $K_{ij}$ and~$I_r$ (cf.\ Eqs.~(\ref{eq:K-Ir1})
and~(\ref{eq:K-Ir2})), with $x_i$ and $-x_i$ replaced by $\sigma_i$ and
$\sigma_i^*$ respectively. Hence 1, $P_{ij}$, $P_{ijk} \equiv P_{ij} P_{jk}$, $\tP$,
$\tilde P_{ij}$, and $\tilde P_{ijk} \equiv P_{ijk} \tilde P$ realize the dihedral
group~$D_6$ in spin space. Such a realization will be referred to as
$D_6^{(s)}$ to distinguish it from the realization $D_6^{(c)}$ in coordinate space,
corresponding to $K_{ij}$ and~$I_r$.\par
%
%
The Hamiltonian ${\cal H}^{(\kappa,\lambda)}$ remains invariant under the combined
action of~$D_6$ in coordinate and spin spaces (to be referred to as $D_6^{(cs)}$),
since it commutes with both $K_{ij} P_{ij}$ and $I_r \tP$. Its eigenfunctions
corresponding to a definite eigenvalue therefore belong to a (reducible or
irreducible) representation of~$D_6^{(cs)}$. For indistinguishable particles that are
bosons (resp.~fermions), only those irreducible representations of $D_6^{(cs)}$ that
contain the symmetric (resp.~antisymmetric) irreducible representation of the
symmetric group $S_3$ should be considered. There are only two such inequivalent
representations, which are both one-dimensional and denoted by $A_1$ and $B_1$
(resp.~$A_2$ and $B_2$)~\cite{hamermesh}. They differ in the eigenvalue of $I_r
\tP$, which is equal to $+1$ or $-1$, respectively.\par
%
%
In such representations, for an appropriate choice of the parameters
$\kappa$,~$\lambda$, ${\cal H}^{(\kappa,\lambda)}$ can be obtained from
$H_{exch}$ by applying some projection operators. Let indeed $\Pi_{B\pm}$
(resp.~$\Pi_{F\pm}$) be the projection operators that consist in replacing $K_{ij}$
and $I_r$ by $P_{ij}$ (resp.~$-P_{ij}$) and $\pm \tP$, respectively, when they
are at the right-hand side of an expression. It is obvious that
$\Pi_{B\pm}(H_{exch}) = {\cal H}^{(\kappa,\pm\lambda)}$, and
$\Pi_{F\pm}(H_{exch}) = {\cal H}^{(-\kappa,\pm\lambda)}$. If $H_{exch}$ has been
diagonalized on a basis of functions depending upon coordinates and spins, then its
eigenfunctions $\Psi(\xb,\sigmab)$ are also eigenfunctions of ${\cal
H}^{(\kappa,\pm\lambda)}$ (resp.~${\cal H}^{(-\kappa,\pm\lambda)}$) provided that
$(K_{ij} - P_{ij}) \Psi(\xb,\sigmab) = 0$ (resp.\ $(K_{ij} + P_{ij}) \Psi(\xb,\sigmab)
= 0$) and $(I_r \mp \tP) \Psi(\xb,\sigmab) = 0$. We shall not pursue the
determination of the eigenfunctions of ${\cal H}^{(\kappa,\lambda)}$ any further,
leaving a detailed derivation for a forthcoming publication.\par
%
%
In conclusion, in the present letter we did propose a three-body generalization of
the Sutherland problem with internal degrees of freedom, related to a
corresponding problem with exchange terms. For the latter, we did construct both 
$D_6$-covariant, but non-commuting, and commuting, but non-$D_6$-covariant
differential-difference operators, in\linebreak terms of which the Hamiltonian can
be expressed in a very simple way. We did show that whereas the former operators
can be derived in a well-defined way from SUSYQM supercharges, there is some
freedom in the choice of the latter.
\par
\newpage
%
%


\begin{thebibliography}{99}

\bibitem{sutherland} B. Sutherland, Phys. Rev. A 4 (1971) 2019; A 5 (1972) 1372;
Phys. Rev. Lett. 34 (1975) 1083.

\bibitem{calogero} F. Calogero, J. Math. Phys. 10 (1969) 2191, 2197; 12 (1971) 419.

\bibitem{cq} C. Quesne, J. Phys. A 28 (1995) 3533.

\bibitem{poly} A. P. Polychronakos, Phys. Rev. Lett. 69 (1992) 703.

\bibitem{perelomov} M. A. Olshanetsky and A. M. Perelomov, Phys. Rep. 94 (1983)
313.

\bibitem{dunkl} C. F. Dunkl, Trans. Am. Math. Soc. 311 (1989) 167.

\bibitem{ha} Z. N. C. Ha and F. D. M. Haldane, Phys. Rev. B 46 (1992) 9359; \\
K. Hikami and M. Wadati, Phys. Lett. A 173 (1993) 263; \\
J. A. Minahan and A. P. Polychronakos, Phys. Lett. B 302 (1993) 265.

\bibitem{bernard} D. Bernard, M. Gaudin, F. D. M. Haldane and V. Pasquier, J. Phys. A
26 (1993) 5219.

\bibitem{buchstaber} V. M. Buchstaber, G. Felder and A. P. Veselov, Elliptic Dunkl
operators, root systems, and functional equations, preprint hep-th/9403178
(1994); \\
T. Yamamoto, Phys. Lett. A 208 (1995) 293.

\bibitem{andrianov} A. A. Andrianov, N. V. Borisov and M. V. Ioffe, Phys. Lett. A
105 (1984) 19; \\
A. A. Andrianov, N. V. Borisov, M. I. Eides and M. V. Ioffe, Phys. Lett. A 109 (1985)
143.

\bibitem{hamermesh} M. Hamermesh, Group theory (Addison-Wesley, Reading, Mass.,
1962).

\bibitem{wolfes} J. Wolfes, J. Math. Phys. 15 (1974) 1420.

\bibitem{marchioro} F. Calogero and C. Marchioro, J. Math. Phys. 15 (1974) 1425.

\end {thebibliography}
\end{document}